\documentclass[aps,amssymb,floatfix,prd,amsmath,preprintnumbers]{revtex4}
\usepackage{epstopdf}
\usepackage{capt-of}
\usepackage{graphicx}  
\usepackage{dcolumn}   
\usepackage{bm}
\begin{document}
\input epsf.tex
\title{\textbf{Dark Energy Cosmological Models with General forms of Scale
Factor}}
\author{B. Mishra\footnote{
Department of Mathematics, Birla Institute of Technology and Science-Pilani,
Hyderabad Campus, Hyderabad-500078, India, email: bivudutta@yahoo.com},
Pratik P. Ray \footnote{
Department of Mathematics, Birla Institute of Technology and Science-Pilani,
Hyderabad Campus, Hyderabad-500078, India, email: pratik.chika9876@gmail.com}
, S.K.J. Pacif \footnote{
Centre of Theoretical Physics, Jamia Millia Islamia, New Delhi-110025,
India, email: shibesh.math@gmail.com} }
\affiliation{ }

\begin{abstract}
\begin{center}
\textbf{Abstract}
\end{center}

In this paper, we have constructed dark energy models in an anisotropic
Bianchi-V space-time and studied the role of anisotropy in the evolution of
dark energy. We have considered anisotropic dark energy fluid with different
pressure gradients along different spatial directions. In order to obtain a
deterministic solution, we have considered three general forms of scale
factor. The different forms of scale factors considered here produce time
varying deceleration parameters in all the cases that simulates the cosmic
transition. The variable equation of state (EoS) parameter, skewness
parameters for all the models are obtained and analyzed. The physical
properties of the models are also discussed.
\end{abstract}

\maketitle

\input epsf.tex 

\input epsf.tex 

\input epsf.tex 


\section{Introduction}

Cosmological constant is the basic choice for studying Dark Energy (DE),
however it has suffered with many difficulties such as the fine tuning and
coincidence problem. Therefore many dynamically varying DE candidates have
been studied \cite{Copel06, Wang16}. The theory of DE is a widely accepted
theory to describe the observed acceleration expansion of the Universe. But
a complete understanding on the nature of dark energy is still remain a challenge to the researchers.
Recent Planck results estimate a maximum contribution of $68\%$ for DE in
the cosmic mass energy budget \cite{Ade14}. Observations have confirmed that
the cosmic speed up is a late time phenomena and has occurred at a redshift
of the order $z_{t}\sim 0.7$. This indicates that, the Universe has
undergone a transition from a decelerated phase of expansion to an
accelerating phase in the recent past. This cosmic transit phenomenon
speculates an evolving deceleration parameter with a signature flipping. The
rate at which the transition occurs usually determines the transit redshift $%
z_{t}$.\newline

The standard cosmological model is based upon the assumption of large scale
isotropy and homogeneity of space. However, a small scale
anisotropies can be expected in the Universe in view of the observations of temperature
anisotropy in the Cosmic Microwave Background (CMB) radiation data from
Wilkinson Microwave Anisotropy Probe (WMAP) and Planck. These data show some
non trivial topology of the large scale cosmic geometry with asymmetric
expansion \citep{ Watan09, Buiny06,tripa2}. Planck data also show a slight
redshit of the power spectrum from exact scale invariance. Though the
standard $\Lambda $CDM model is well supported by different measurements, at
least at low multipoles, it does not fit to the temperature power spectrum
data \cite{Ade14}. Dark energy is also believed to be associated with the
breakdown of global isotropy and can display anisotropic features \cite%
{Picon04, Jaffe05, Koivisto08a, Cooray10}. Employing the Union compilation
of Type Ia Supernovae, Cook and Lynden-Bell \cite{Cook10} found that a weak anisotropy in
dark energy, mostly observable for higher redshift group with $z>0.56$,
directed roughly towards the cosmic microwave background dipole.
While Friedmann Robertson Walker models (FRW) in the framework of General
Relativity (GR) provide satisfactory explanations to various issues in
cosmology, Bianchi type models have been constructed in recent times to
handle the smallness in the angular power spectrum of temperature anisotropy 
\cite{campa1, campa3, gruppo, Koivisto06, Koivisto08, Bat09, akarsu1,
akarsu2}. Moreover, Bianchi type models are the generalisation of the FRW
models, where the spatial isotropy is relaxed.\\

As a sequel to our previous studies on the dynamical behaviour of pressure
anisotropies \cite{Mishra15, mishrampla, skt15}, in the present work, we
have considered different general form of non interacting dark energy with
anisotropic pressures along different spatial directions and have
investigated the effect of anisotropic components on the dynamics of dark
energy. Most recently Mishra et al. \cite{Mishra17a, Mishra17b} have
investigated the anisotropic behaviour of the accelerating universe in
Bianchi V space time in the frame work of General Relativity (GR), when the
matter field is considered to be of two fluids the usual (a) geometric
string with DE and (b) nambu string with DE fluid. The present paper is organized as
follows. In Sect. 2, we discuss the basic formalism of an anisotropic dark
energy in a Bianchi type V space time. The effect of anisotropy on dark energy and the DE equation of state (EoS) parameter along with the behaviour of other physical parameters are discussed in Sect. 3 and we summarize our results and its discussions in Sect. 4.

\section{Basic Formalism of the Model and Anisotropy}

We consider the line element as Bianchi V (BV) space time in the form

\begin{equation}  \label{eq:1}
ds^{2}= -dt^{2}+a_1^2 dx^2+e^{2kx}(a_2^2 dy^2+a_3^2 dz^2)
\end{equation}
Here, $a_i$; $i=1,2,3$ are the metric potentials, which are functions of
time $t$ only, The metric potentials $a_i$'s are assumed to be different in
different directions. So, it provides a description for anisotropic
expansions along three orthogonal directions. When all the metric potentials
are same, the model reduces to FRW model. Here, $k$ is a non zero arbitrary
constant and $8\pi G=c=1$, $G$ is the Newtonian gravitational constant and $%
c $ is the speed of light. We have defined the energy momentum tensor for
Dark Energy (DE) as

\begin{equation}
T_{\mu \nu }=(\rho +p)u_{\mu }u_{\nu }+pg_{\mu \nu }  \label{eq:2}
\end{equation}%
where $u^{\mu }u_{\mu }=1$ and in a co-moving coordinate system, $u^{\mu }$
is the four velocity vector of the fluid. $p$ and $\rho $ are respectively the isotropic
pressure of the fluid and the proper energy density. We are interested to
incorporate some amount of anisotropy in the DE pressure, therefore the
energy momentum tensor can be rewritten as:

\begin{eqnarray}  \label{eq:3}
T_{\mu\nu} & = & diag[\rho, -p_x,-p_y,-p_z]  \nonumber \\
& = & diag[1, -\omega_x,-\omega_y,-\omega_z]\rho \\
& = & diag[1, -(\omega+\delta),-(\omega+\gamma),-(\omega+\eta)]\rho, 
\nonumber
\end{eqnarray}
where $\rho$ and $\omega$ respectively denote the density and equation of
state (EoS) parameter of DE. $\delta$, $\gamma$ and $\eta$ are the skewness
parameters which deviates from the EoS parameter along $x$, $y$ and $z$ axes
respectively. When the skewness parameters vanish identically the pressure
of DE becomes isotropic. Now, the field equations, $R_{\mu\nu}-\frac{1}{2}%
Rg_{\mu \nu}=-T_{\mu\nu}$, for BV space time in the frame work of GR when
the matter field is in the form of DE can be obtained as

\begin{equation}  \label{eq:4}
\frac{\ddot{a_2}}{a_2}+\frac{\ddot{a_3}}{a_3}+\frac{\dot{a_2}\dot{a_3}}{%
a_2a_3}-\frac{k^{2}}{a_1^{2}}=-(\omega+\delta)\rho
\end{equation}

\begin{equation}  \label{eq:5}
\frac{\ddot{a_1}}{a_1}+\frac{\ddot{a_3}}{a_3}+\frac{\dot{a_1}\dot{a_3}}{%
a_1a_3}-\frac{k^{2}}{a_1^{2}}=-(\omega+\gamma)\rho
\end{equation}

\begin{equation}  \label{eq:6}
\frac{\ddot{a_1}}{a_1}+\frac{\ddot{a_2}}{a_2}+\frac{\dot{a_1}\dot{a_2}}{%
a_1a_2}-\frac{k^{2}}{a_1^{2}}=-(\omega+\eta)\rho
\end{equation}

\begin{equation}  \label{eq:7}
\frac{\dot{a_1}\dot{a_2}}{a_1a_2}+\frac{\dot{a_2}\dot{a_3}}{a_2a_3}+\frac{%
\dot{a_3}\dot{a_1}}{a_3a_1}-\frac{3k^{2}}{a_1^{2}}=\rho
\end{equation}

\begin{equation}  \label{eq:8}
2\frac{\dot{a_{1}}}{a_{1}}-\frac{\dot{a_{2}}}{a_{2}}-\frac{\dot{a_{3}}}{a_{3}%
}=0,
\end{equation}
where an over dot represents the differentiation with respect to the cosmic
time $t$. By suitably absorbing the integrating constants in the metric
potentials, from eqn. \eqref{eq:8}, we obtain $a_{1}^{2}=a_{2}a_{3}$. To
obtain an anistropic relation among the directional scale factors, here we
have assumed $a_{2}=a_{3}^{l}$ where $l$ is an arbitrary positive constant $%
\neq 1$.\newline

We have assumed the rate of expansion in $x$-direction and the average
expansion are same. The mean Hubble parameter $H$ and directional Hubble
parameters $H_{i},i=1,2,3$ along different spatial directions can be
expressed respectively as $H=\frac{\dot{R}}{R}=\frac{1}{3}%
\sum_{i=1}^{3}H_{i} $ and $H_{i}=\frac{\dot{a_{i}}}{a_{i}}$. Also the cosmic
spatial volume $V$, scalar expansion $\theta $ and shear scalar $\sigma ^{2}$
can be defined respectively as $V=R^{3}=a_{1}a_{2}a_{3}$, $\theta =3H$, and $%
\sigma ^{2}=\frac{1}{2}(\sum_{i=1}^{3}H_{i}^{2}-\frac{\theta ^{2}}{3})$.
Moreover, the deceleration parameter can be obtained from the relation $q=%
\frac{d}{dt}\left( \frac{1}{H}\right) -1$. With the anisotropy assumptions $%
a_{2}=a_{3}^{l}$, the scale factors along different directions can be $%
a_{1}=R,a_{2}=R^{\frac{2l}{l+1}}$ and $a_{3}=R^{\frac{2}{l+1}}$.
Subsequently, the directional Hubble parameters can be obtained as, $H_{1}=H$%
, $H_{2}=\big(\frac{2l}{l+1}\big)H$ and $H_{3}=\big(\frac{2}{l+1}\big)H$. We
can also obtain the anisotropic parameter $\mathcal{A}$ as ${\mathcal{A}}=%
\frac{1}{3}\sum \left( 1-\frac{H_{i}}{H}\right) ^{2}=\frac{2}{3}\left( \frac{%
l-1}{l+1}\right) ^{2}$. For $l=1$, the anisotropy parameter $\mathcal{A}$
vanishes implying the model becomes isotropic with the rate of expansion is
equal in all spatial directions. The energy conservation equation for the
anisotropic fluid, $T_{;\nu }^{\mu \nu }=0$, which will subsequently split
into two parts, can be defined as

\begin{equation}  \label{eq:9}
\dot{\rho}+3\rho(\omega+1)H+\rho(\delta H_1+\gamma H_2+\eta H_3)=0.
\end{equation}
In the above equation, when the third term is zero, it reduces to eqn.%
\eqref{eq:10}, which defines the conservation of matter field with equal
pressure in all directions whereas when the first two terms are zero, eqn. %
\eqref{eq:9} reduces to eqn.\eqref{eq:11}, which is the deviations of
skewness parameters only.

\begin{equation}  \label{eq:10}
\dot{\rho}+3H(\omega +1)\rho=0
\end{equation}
and 
\begin{equation}  \label{eq:11}
\dot{\rho}+ \rho(\delta H_1+\gamma H_2+\eta H_3)=0.
\end{equation}

Using the mathematical scheme developed in our earlier work, \cite%
{mishrampla}, the skewness parameters can be obtained as

\begin{eqnarray}
\delta &=&-\chi (l)\left( \frac{l-1}{\rho }\right) f(H)  \label{eq:12} \\
\gamma &=&\chi (l)\left( \frac{l+5}{2\rho }\right) f(H)  \label{eq:13} \\
\eta &=&-\chi (l)\left( \frac{5l+1}{2\rho }\right) f(H)  \label{eq:14}
\end{eqnarray}

where $\chi (l)= \dfrac{l-1}{(l+1)^2}$, represents the amount of deviation
from isotropic behaviour of the model and from the parameter $l$. The
pressure anisotropies are also decided from the behaviour of the function $%
f(H)=\frac{2}{3}(\dot{H}+ 3H^2)$. The energy density and EoS parameter can be respectively obtained as

\begin{equation}  \label{eq:15}
\rho =2\left[ \frac{l^{2}+4l+1}{(l+1)^{2}}\right] H^{2}-3\frac{k^{2}}{R^{2}}
\end{equation}

\begin{equation}  \label{eq:16}
\omega =-\frac{2}{\rho }\left[ \frac{l^{2}+4l+1}{(l+1)^{2}}\right] \left[
f(H)-H^{2}\right] +\frac{k^{2}}{\rho R^{2}}
\end{equation}

We can observe from \eqref{eq:15}, the positivity condition of energy
density $\rho $ requires $2\left[ \frac{l^{2}+4l+1}{(l+1)^{2}}\right] H^{2}-3%
\frac{k^{2}}{R^{2}}>0$ implying that at any time interval we should have 
\begin{equation}
H^{2}>1.5k^{2}\frac{(l+1)^{2}}{l^{2}+4l+1}\frac{1}{R^{2}}\text{.}  \label{17}
\end{equation}

This general condition i.e. $H^{2}>\frac{constant}{R^{2}}$ should be
satisfied while studying any models of Bianchi-V space-time. The most
general functional form of the scale factor $R$ satisfying this condition is
an exponent function. In this paper in order to study the background
cosmology, we will consider three functional forms of scale factor
containing an exponent term.

\section{Scale Factors and Anisotropy Effect}

The variation of Hubble law may not be consistent with observations.
However, it has the advantage of providing simple functional form of the
time evolution of the scale factor. Also in order to explain the late time cosmic acceleration a simple parametrization of Hubble parameter is needed \cite{akarsu3,Pacif16}.  Hence, in order to construct the
cosmological models of the universe, in this section, we have adopted three
general forms of the scale factor $R$.

\subsection{Case I}

In this case, we have considered the scale factor which is a combination of two factors:
the one is the usual power law expansion and the second one is the
generalized form of an exponential function. Another reason to consider this
scale factor is that a cosmic transit from early deceleration to late time
acceleration can be obtained in this form of scale factor. So, the scale factor can be defined as 
\begin{equation}
R(t)=t^m\xi ^{n t}  \label{eq:18}
\end{equation}

where $m$, $n$, $\xi $ are positive constants and $2\leq \xi \leq 3$. For $%
\xi =2.718$ (Exponential), it reduces to hybrid scale factor cosmology \cite%
{mishrampla}. So, the Hubble parameter becomes $nln\xi +\frac{m}{t}$.
Subsequently the directional Hubble parameters can be obtained as: $%
H_{1}=H=nln\xi +\frac{m}{t}$, $H_{2}=\left( \frac{2l}{l+1}\right) \left(
nln\xi +\frac{m}{t}\right) $ and $H_{3}=\left( \frac{2}{l+1}\right) \left(
nln\xi +\frac{m}{t}\right) $. Now, we can obtain the functional $f(H)$, for
the considered Hubble parameter as 
\begin{equation}
f(t)=\frac{2}{3}\left[ 3\left( nln\xi +\frac{m}{t}\right) ^{2}-\frac{m}{t^{2}%
}\right]  \label{eq:19}
\end{equation}

So, the energy density \eqref{eq:15} and the EoS parameter \eqref{eq:16} can
be further simplified as:

\begin{equation}  \label{eq:20}
\rho=2 \left(\frac{l^2+4l+1}{l^2+2l+1}\right) \left(n ln\xi +\frac{m}{t}%
\right)^2 -3\left(\frac{k}{t^{m}\xi ^{nt}}\right)^2
\end{equation}

\begin{equation}
\omega =-\frac{2}{\rho }\left( \frac{l^{2}+4l+1}{l^{2}+2l+1}\right) \left[
\left( nln\xi +\frac{m}{t}\right) ^{2}-\frac{2m}{3t^{2}}\right] +\frac{1}{\rho}\left( 
\frac{k}{t^{m}\xi ^{nt}}\right) ^{2}  \label{eq:21}
\end{equation}

\begin{figure}[tbph]
\minipage{0.40\textwidth}
\centering
\includegraphics[width=\textwidth]{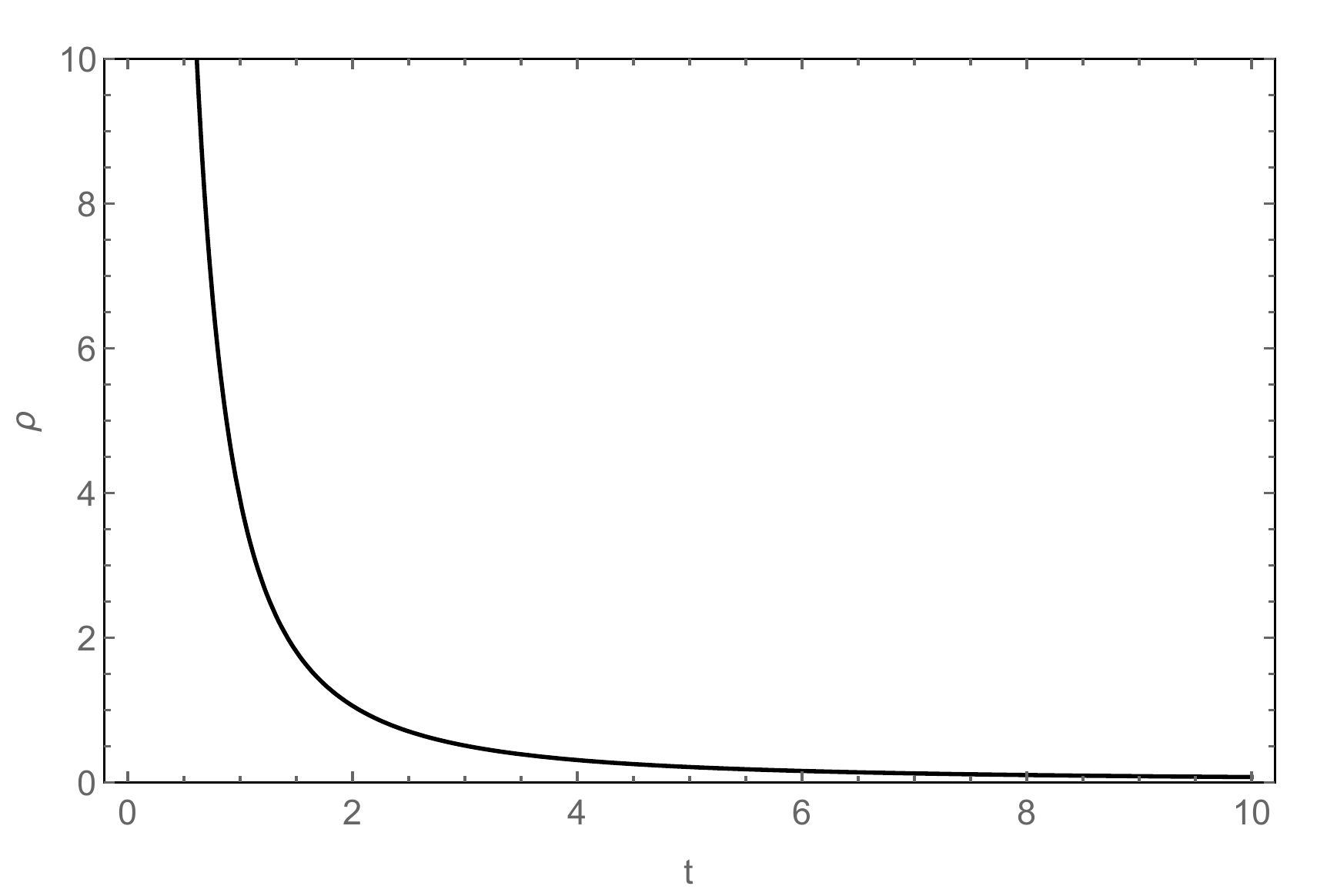}
\caption{ Energy density $\protect\rho$ versus time $t$ \\ 
($m=1.1,$ $n=0.05,$ $k=0.1,$ $l=0.9$)}
\endminipage
\minipage{0.40\textwidth}
\centering
\includegraphics[width=\textwidth]{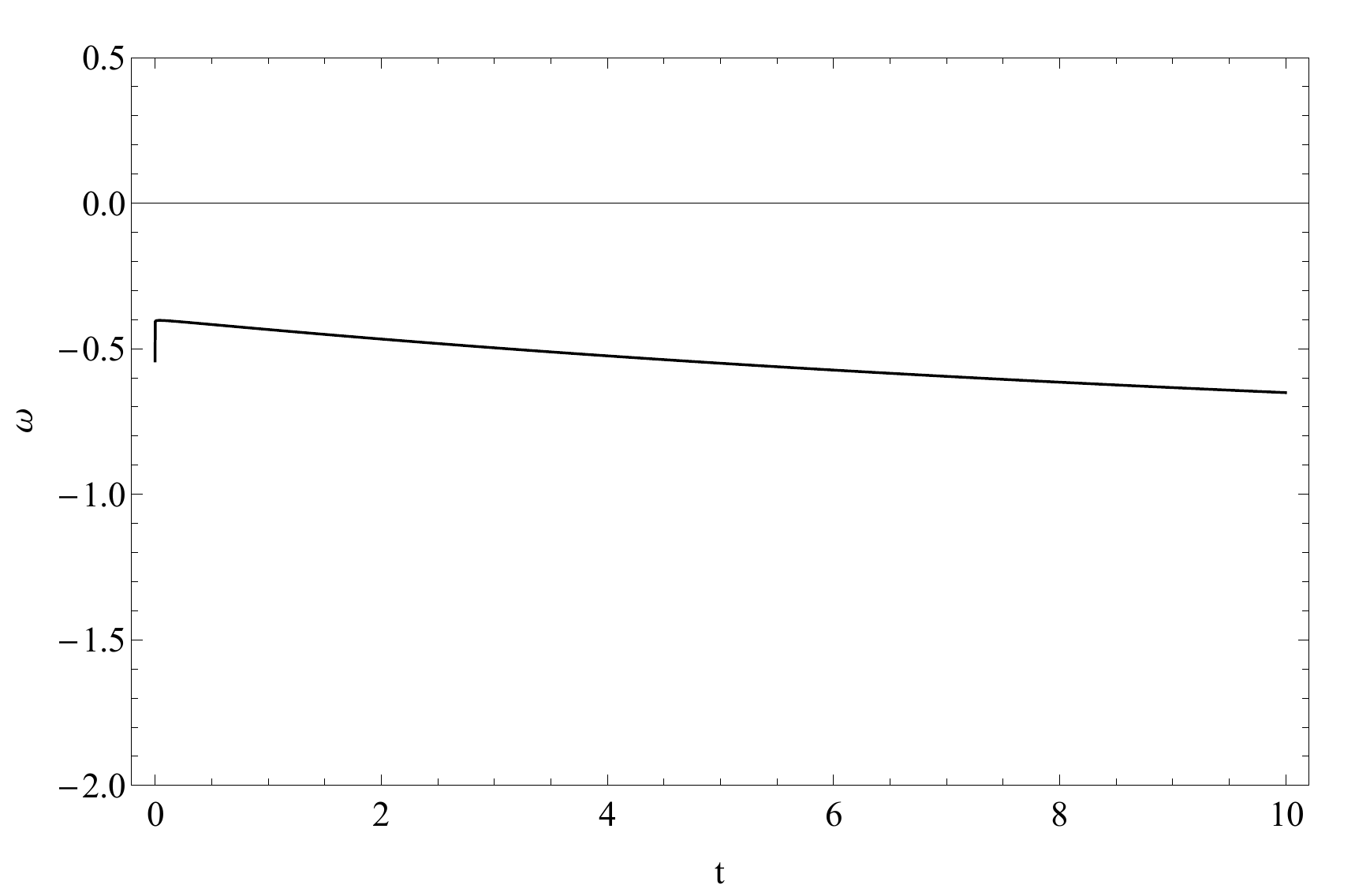}
\caption{ EoS parameter $\protect\omega$ versus time $t$\\
($m=1.1,$ $n=0.05,$ $k=0.1,$ $l=0.9$)}
\endminipage
\end{figure}

where $\chi(l)=\frac{l-1}{(l+1)^2}$.The plots for the energy density and EoS parameter has been shown in FIG-1
and FIG-2 respectively by choosing appropriate adjustable constants to see
the time evolution of these parameters. The behaviour of the energy density
would be decided by the power of the cosmic time in the expression %
\eqref{eq:20}. At the late cosmic
phase the energy density $\rho \simeq 2\left( \frac{l^{2}+4l+1}{l^{2}+2l+1}%
\right) \left( nln\xi \right) ^{2}$. It has been observed that both in early
time and late time the energy density remain positive. It might be negative
if the second term in eqn. \eqref{eq:20} dominates the first term. However,
we have already adjusted the functions in such a way that the energy density
remains positive throughout the cosmic evolution. The EoS parameter starts in the negative region and stays in the acceptable range \cite{Mathew15}. Hence, it can be infer that the matter evolves in the quintessence region at early time and remains in the same domain till late phase of cosmic time and approaching towards phantom barrier. The parameter evolves
dynamically with the expansion of the universe. This is basically governed
by the rest energy density that appears in the denominator of eqn. \eqref{eq:21}.
Moreover, it is worthy to mention here that since the exponent $n$ and the
value of $\xi $ are positive, at late phase $\omega \sim -1$. Subsequently,
we obtained the skewness parameters as:

\begin{equation}  \label{eq:22}
\delta = -\left(\frac{l-1}{\rho}\right)\left(\frac{2}{3}\chi(l)\right)\left[%
3\left(n ln\xi +\frac{m}{t}\right)^2-\frac{m}{t^2}\right]
\end{equation}

\begin{equation}  \label{eq:23}
\gamma = \left(\frac{l+5}{2\rho}\right)\left(\frac{2}{3}\chi(l)\right)\left[%
3\left(n ln\xi +\frac{m}{t}\right)^2-\frac{m}{t^2}\right]
\end{equation}

\begin{equation}  \label{eq:24}
\eta = -\left(\frac{5l+1}{2\rho}\right)\left(\frac{2}{3}\chi(l)\right)\left[%
3\left(n ln\xi +\frac{m}{t}\right)^2-\frac{m}{t^2}\right]
\end{equation}

\begin{figure}[tbph]
\minipage{0.40\textwidth}
\centering
\includegraphics[width=\textwidth]{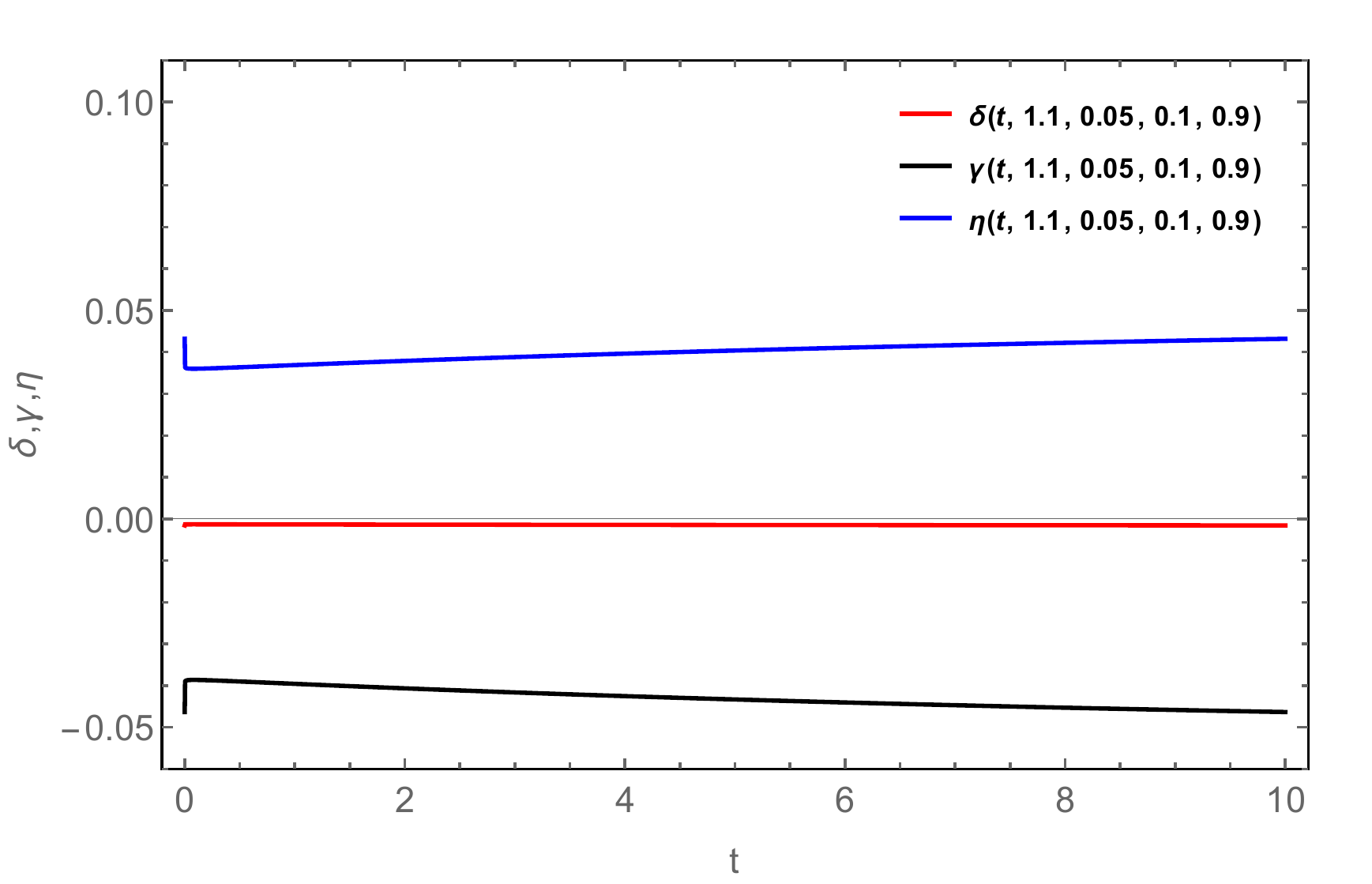} 
\caption{Skewness parameters versus time $t$\\
($m=1.1,$ $n=0.05,$ $k=0.1,$ $l=0.9)$}
\endminipage
\label{fig:case-1-DGE}
\minipage{0.40\textwidth}
\centering
\includegraphics[width=\textwidth]{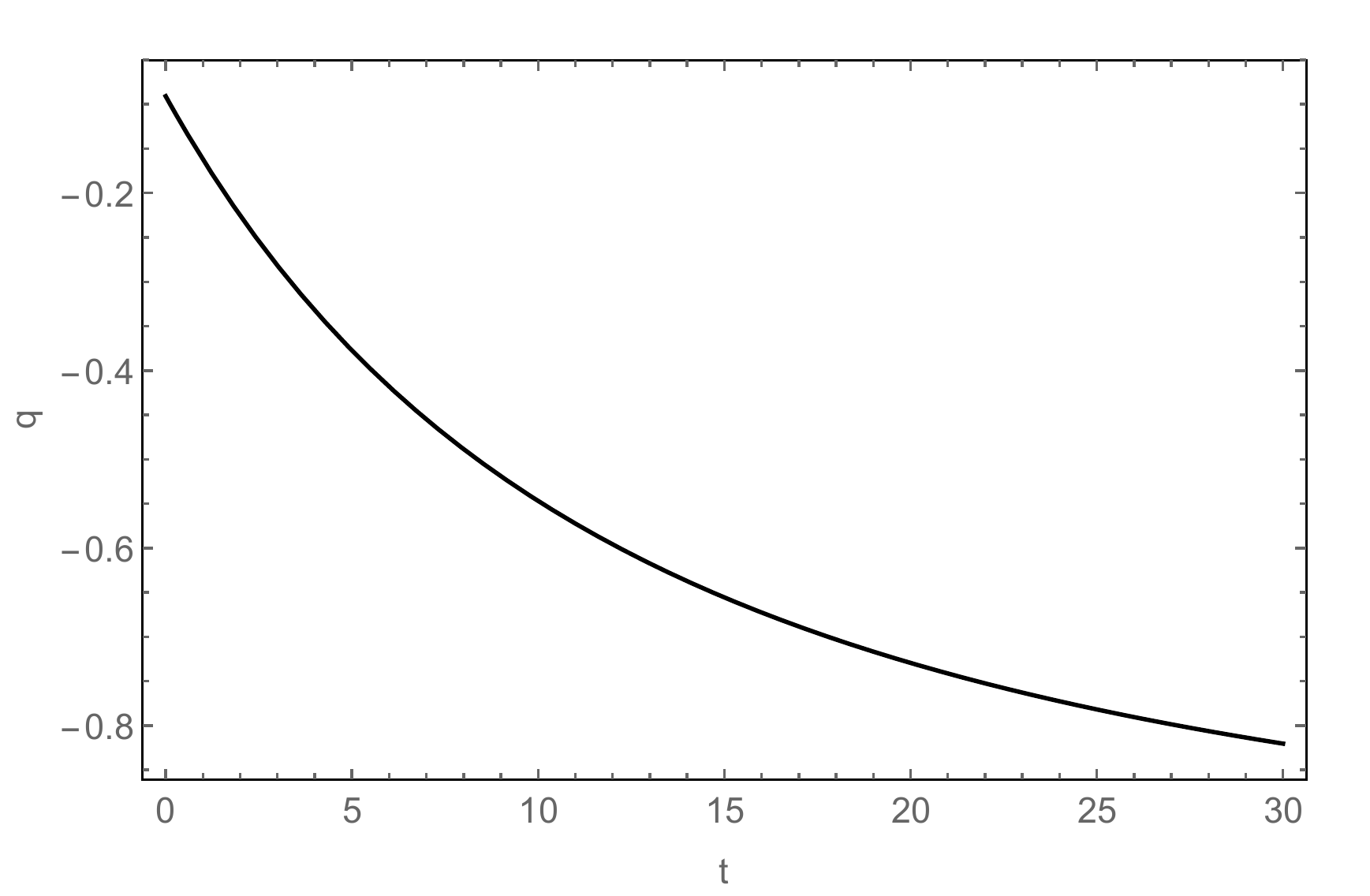}
\caption{Deceleration parameter $q$ versus time $t$\\
($m=1.1,$ $n=0.05,$ $k=0.1,$ $l=0.9$)}
\endminipage
\end{figure}

The plot for the anisotropic parameters and cosmic time has been represented
in FIG-3. The functional $\frac {f(t)}{\rho}$ is also time varying, as the
energy density $\rho$ has been changing with the time. So, the evolutionary
behaviour of the skewness parameters $\delta$, $\eta$ and $\gamma$ which are
negative at early cosmic time and positive in the late phase have been
decided by the factor $\frac {f(t)}{\rho}$. As expected, the skewness
parameter along $x-$axis does not evolve with time and mostly remains close
to zero. This expectation is because in our initial assumption along $x-$%
axis, the expansion rate is assumed to be the same as the mean Hubble rate.
It also indicates that the the pressure of the anisotropic fluid also same
with the mean pressure. It is also very much notable that the other skewness
parameters $\gamma$ and $\eta$ are found to be the mirror image of each
other. This may be due to the fact that the anisotropic parameter $l$ is
close to 1, which will describe the small anisotropic nature of the universe (FIG-4). The constants (model parameters) are adjusted in such a way that the present numerical values of the cosmological parameters lie in the neighbourhood of predicted values by the observational data. In FIG-4, $q(t_0)\sim-0.8$ where $t_0=13.7$ billion years. Here, 2 units of cosmic time $t$ measured as 1 billion year. So, the current assigned value for cosmic time $t_0$ = 28 units, gives $t_0\sim13.7$ billion years.\\

The scalar expansion and shear scalar of the model can be obtained
respectively as $\theta=3n ln\xi+\frac{3m}{t}$ and $\sigma^2=\left(\frac{l-1%
}{l+1}\right)^2\left(nln\xi+\frac{m}{t}\right)^2$. These parameters confirm the rate of change of local volume of the corresponding model in an increasing way since all $m$,$n$ and $\xi$ are positive. Also, the deceleration parameter can be obtained as $q=-\frac{1}{H^2}\left((nln\xi)^2+\frac{m(m-1)}{%
t^2}+\frac{m}{t}\right)$. The deceleration parameter remains negative and time dependent for $m>1$ and $n>0$ with a negative value at early phase and $q\sim -1$ at late times of acceleration agreeing with the present observational result.

\subsection{Case II}

In this case, we have considered the scale factor as a hyperbolic function
in the form 
\begin{equation}  \label{eq:25 }
R(t)=\alpha cosh \beta t
\end{equation}
where $\alpha $, $\beta $ are positive constants and the Hubble parameter
obtained as $H=\beta tanh\beta t$. Subsequently the directional Hubble
parameters can be obtained as: $H_{1}=H=\left( \beta tanh\beta t\right) $, $%
H_{2}=\left( \frac{2l}{l+1}\right) \left( \beta tanh\beta t\right) $ and $%
H_{3}=\left( \frac{2}{l+1}\right) \left( \beta tanh\beta t\right) $. Now, we
can obtain the functional $f(H)$, for the considered Hubble parameter as 
\begin{equation}
f(t)=\frac{2}{3}\left[ \beta ^{2}\left( 1+2tanh^{2}\beta t\right) \right]
\label{eq:26}
\end{equation}

The energy density and the EoS parameter can be expressed as:

\begin{equation}  \label{eq:27}
\rho=2 \left(\frac{l^2+4l+1}{l^2+2l+1}\right) \left(\beta tanh \beta
t\right)^2 -3\left(\frac{k}{\alpha cosh \beta t}\right)^2
\end{equation}

\begin{equation}  \label{eq:28}
\omega=-\frac{2}{\rho}\left(\frac{l^2+4l+1}{l^2+2l+1}\right) \left[\frac{2}{3%
}\beta^2\left(1+\frac{1}{3} tanh^2\beta t\right)\right]+\frac{1}{\rho}\left(\frac{k}{%
\alpha cosh \beta t}\right)^2
\end{equation}

\begin{figure}[tbph]
\minipage{0.40\textwidth} 
\centering
\includegraphics[width=\textwidth]{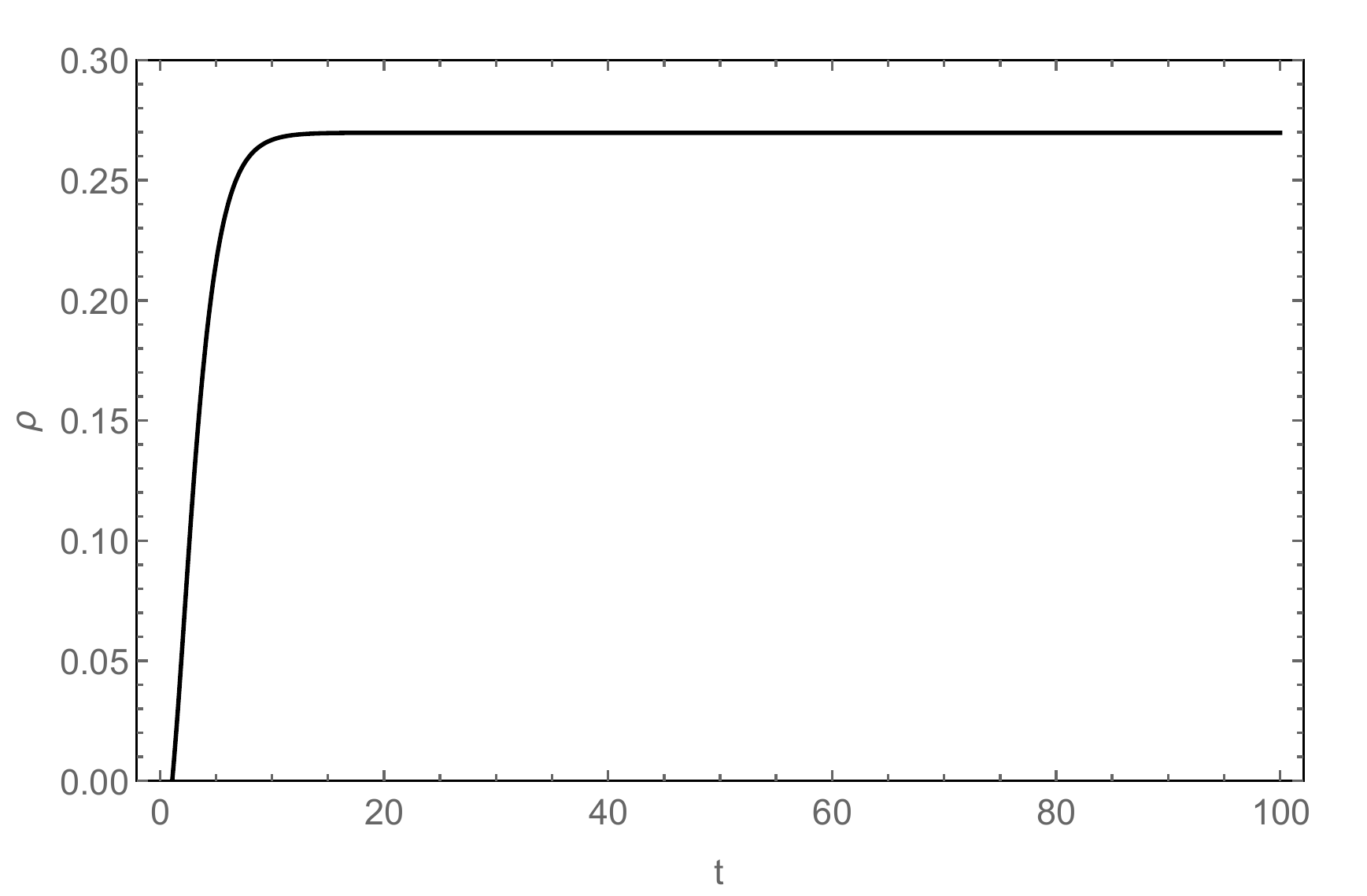}
\caption{ Energy density $\protect\rho$ versus cosmic time $t$ \newline
($\protect\alpha =1,$ $\protect\beta =0.30,$ $k=0.1,$ $l=0.9)$}
\endminipage
\minipage{0.40\textwidth}
\centering
\includegraphics[width=\textwidth]{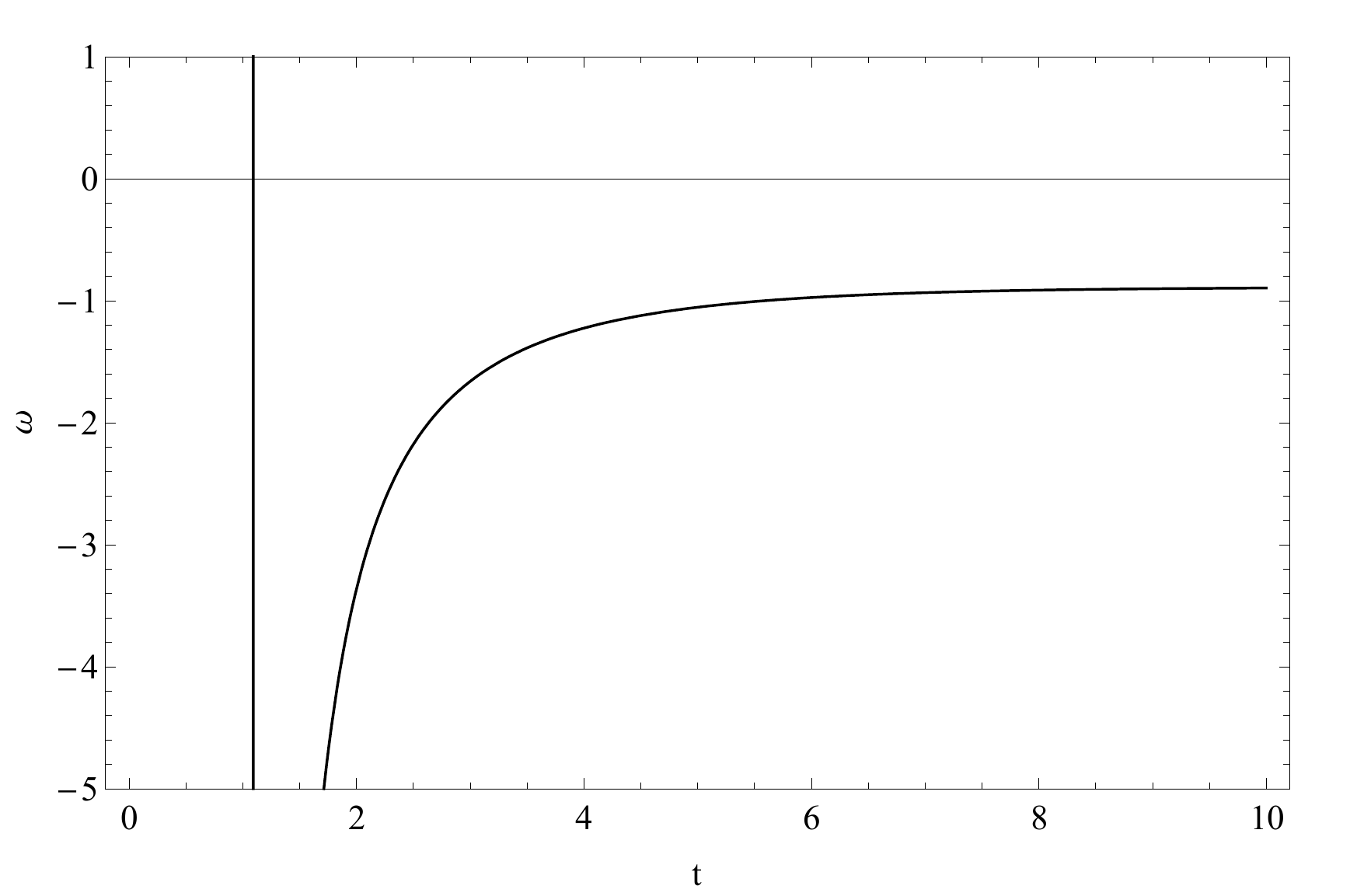} 
\caption{ EoS parameter $\protect\omega$ versus cosmic time $t$ \newline
($\protect\alpha =1,$ $\protect\beta =0.30,$ $k=0.1,$ $l=0.9)$}
\endminipage


\end{figure}

It has been observed that the energy density increases with increase in time and remain flat after a
particular time period. Initially there is a sharp increase and subsequently
the increment becomes constant at late time. The details has been shown in
FIG-5. It may be noted here that since $\rho$ needs to be positive, the
first term of \eqref{eq:27}, should dominates the second. Therefore the
behaviours of the parameters are constrained accordingly within the
admissible limit. The EoS parameter in FIG-6 shows phantom behaviour in early phase of cosmic time ($t\sim 1.7 $ billion years) and evolves dynamically. However, it crosses the phantom barrier at $t=5.8$ before stabilising in the quintessence region as referred by current Planck satellite data ( $\omega=-1.49_{-0.57} ^{+0.05}$ )\cite{Mathew15, Brevik15}. The skewness parameters of the model with
this scale factor can be obtained as:

\begin{equation}  \label{eq:29}
\delta = -\left(\frac{l-1}{\rho}\right) \left(\frac{2}{3}\chi(l) \right)%
\left[\beta^2\left(1+2tanh^2\beta t\right)\right]
\end{equation}

\begin{equation}  \label{eq:30}
\gamma = \left(\frac{l+5}{2\rho}\right)\left(\frac{2}{3}\chi(l)\right)\left[%
\beta^2\left(1+2 tanh^2\beta t\right)\right]
\end{equation}

\begin{equation}  \label{eq:31}
\eta = -\left(\frac{5l+1}{2\rho}\right)\left(\frac{2}{3}\chi(l)\right)\left[%
\beta^2\left(1+2 tanh^2\beta t\right)\right]
\end{equation}

\begin{figure}[tbph]
\minipage{0.40\textwidth}
\centering
\includegraphics[width=\textwidth]{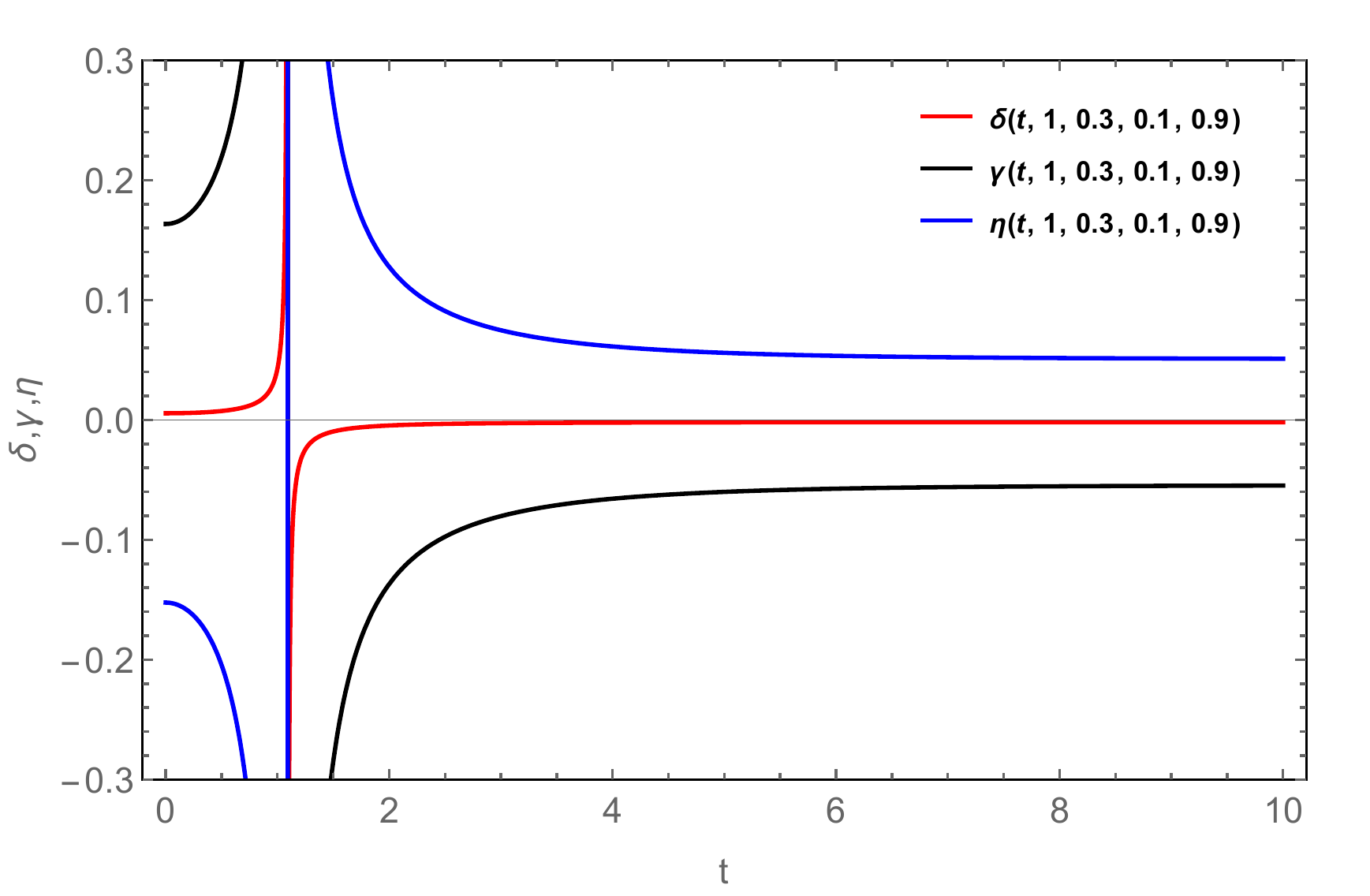}
\caption{Skewness parameters versus time $t$\\
($\protect\alpha =1,$ $\protect\beta =0.30,$ $k=0.1,$ $l=0.9)$}
\endminipage
\minipage{0.40\textwidth}
\centering
\includegraphics[width=\textwidth]{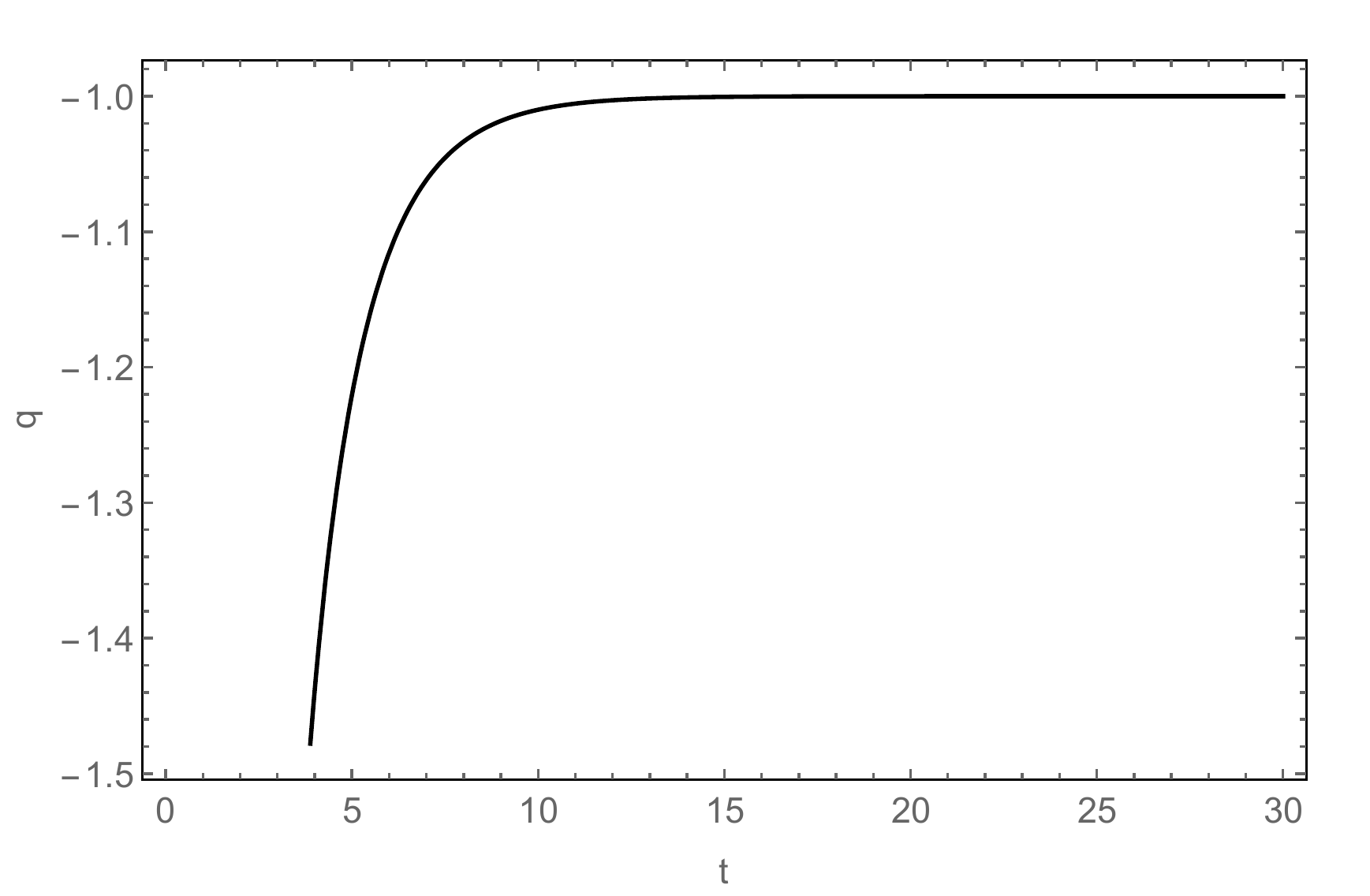}
\caption{Deceleration parameter $q$ versus time $t$
($\protect\alpha =1,$ $\protect\beta =0.30,$ $k=0.1,$ $l=0.9)$}
\endminipage
\end{figure}

In FIG-7, we have represented the behaviour of skewness parameters with respect
to the cosmic time. As in the previous case, in this case also the skewness
parameter $\delta$ does not evolve with time and move all along the x-axis
with increase in $t$. In fact it lies entirely on the $x-$axis at late
times. At the same time the other two skewness parameters $\gamma$ and $\eta$
remain constant at late times respectively in the positive and negative side
of $x-$axis. Both these parameters maintain a uniform distance from the
parameter $\delta$. The scalar expansion of the model can be calculated as $%
\theta=3\beta tanh\beta t$, which indicates that the expansion increases
with increase in time. The shear scalar found to be  $\sigma^2=\left(\frac{l-1}{l+1}%
\right)^2 \beta^2 tanh^2 \beta t$. Irrespective of the value of the
aniostropic value $l$, the shear scalar is positive through out the
evolution. Both the shear scalar and scalar expansion indicate that the rate of deformation is positive with respect to the cosmic evolution through the transition. Scalar expansion remains positive for positive value of $\beta$ whereas in shear scalar $l$ need to constraint as $l>1$ in order to get a positive cosmic behaviour. The deceleration parameter is found to be $q=-tanh^2\beta t$, it
indicates that the parameter always remain negative and close to $-1$ for $\beta=0.30$, adhere to a good match with the observational data. It is also observed from FIG-8 that the accelerated expansion occur in a reasonable time period, which can be termed as the transition period.

\subsection{Case III}

In the previous two cases, we have constructed the dark energy models by
considering the scale factor as a hybrid and hyperbolic functions. In order
to have a better understanding on the anisotropic universe, in this case, we
have considered the scale factor as a fraction of exponential function in
the form:

\begin{equation}  \label{eq:32}
H(R)=a(R^{-n}+1)=\frac{ae^{nat}}{e^{nat}-1}
\end{equation}

where $a$, $b$ are positive constants. The scale factor $R$ can be expressed
as $R=(e^{nat}-1)^{\frac{1}{n}}$. Now, we can obtain the functional $f(H)$
as 
\begin{equation}
f(t)=\frac{2}{3}\left[ -na^{2}e^{nat}+\frac{3a^2e^{2nat}}{(e^{nat}-1)^{2}}%
\right]  \label{eq:33}
\end{equation}

The energy density and EoS parameter can be obtained as,

\begin{equation}  \label{eq:34}
\rho=2 \left(\frac{l^2+4l+1}{l^2+2l+1}\right) \left(\frac{ae^{nat}}{e^{nat}-1%
}\right)^2 -3\left(\frac{k}{(e^{nat}-1)^\frac{1}{n}}\right)^2
\end{equation}

\begin{equation}  \label{eq:35}
\omega =-\frac{2}{\rho }\left( \frac{l^{2}+4l+1}{l^{2}+2l+1}\right) \left[ -%
\frac{2na^{2}e^{nat}}{3}+\left( \frac{ae^{nat}}{e^{nat}-1}\right) ^{2}\right]
+\frac{1}{\rho}\left( \frac{k}{\left( e^{nat}-1\right) ^{\frac{1}{n}}}\right) ^{2}
\end{equation}

\begin{figure}[tbph]
\minipage{0.40\textwidth}
\centering
\includegraphics[width=\textwidth]{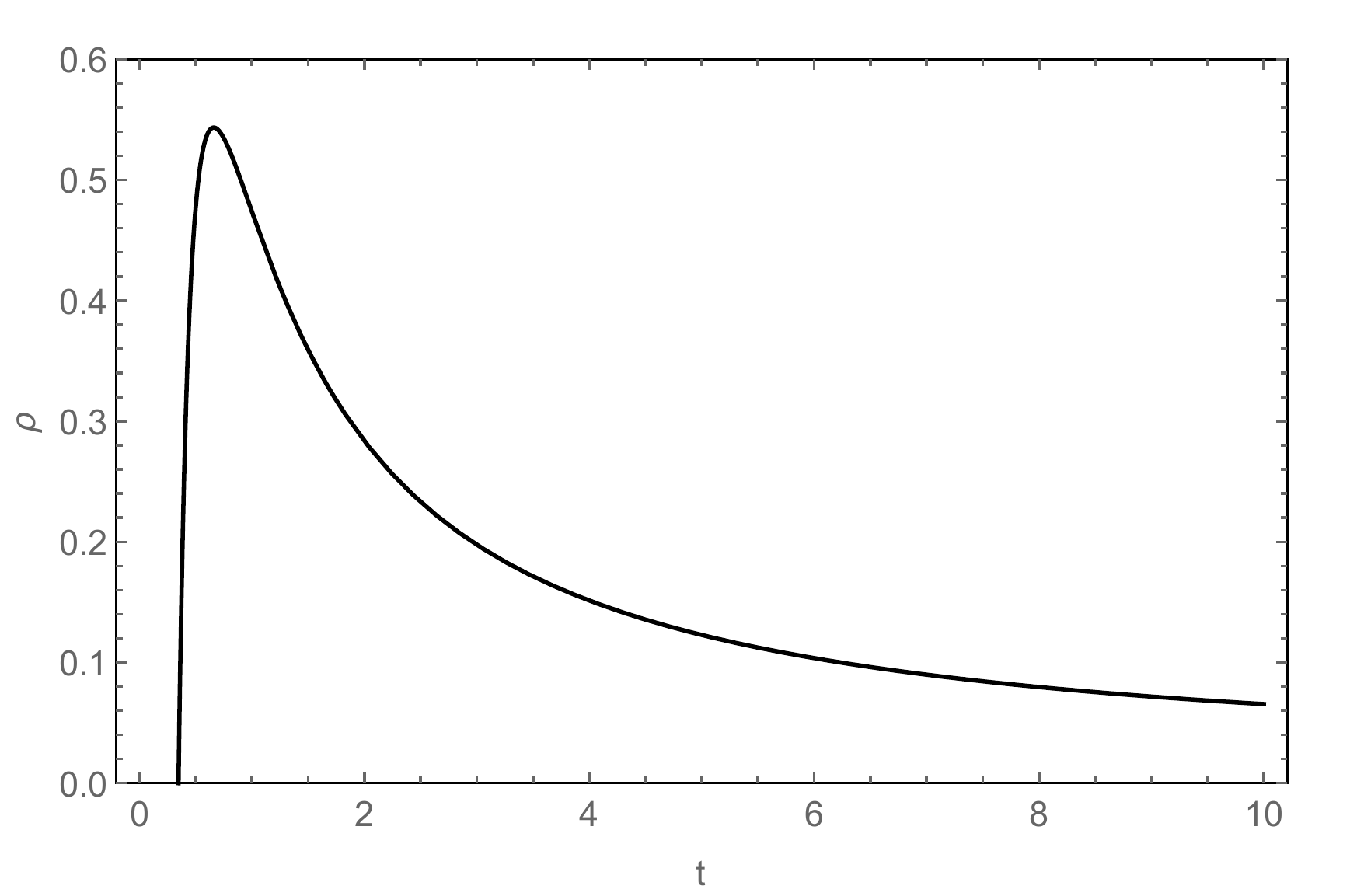} 
\caption{ Energy density $\protect\rho$ versus time $t$ \newline
($n=0.99,$ $a=0.1,$ $k=0.1,$ $l=0.9)$}
\endminipage
\minipage{0.40\textwidth} 
\centering
\includegraphics[width=\textwidth]{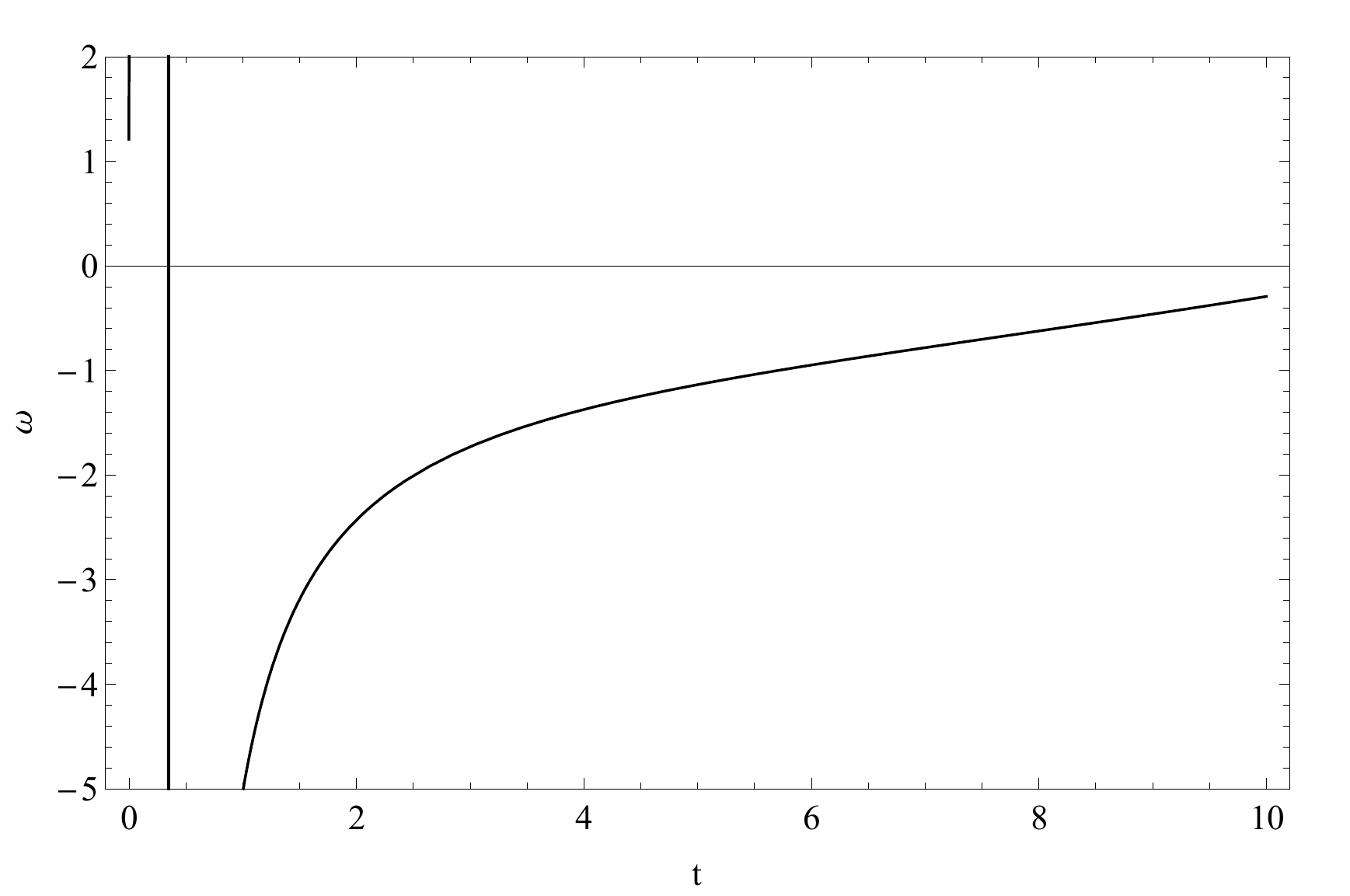} 
\caption{ EoS parameter $\protect\omega$ versus time $t$ \newline
($n=0.99,$ $a=0.1,$ $k=0.1,$ $l=0.9)$}
\endminipage

\end{figure}

The energy density in FIG-9 lies in the positive domain. Though initially, it
rises sharply but at late times it decreases rapidly. The most common type
of dark energy is the EoS parameter with the ratio of pressure and dark
energy density is a constant. Such an EoS is well known if $\omega \in[0,1]$%
, which describes a perfect fluid; however with $\omega=-1$, it defines a
cosmological constant. When $\omega <-\frac{1}{3}$, the universe is filled
with some similar substance which leads to the expansion with acceleration.
However, in most of the cases it remains in the range $({-\frac{1}{3},-1})$. From FIG-10, it
is observed that, $\omega$ evolves in the phantom era at early cosmic time and slides through the quintessence region at late time, crossing the phantom barrier at $t=5.75$ billion years. The parameter proceeds towards still higher values in later evolution and remains in the preferred observed range. The skewness
parameters for the model can be obtained as:

\begin{equation}  \label{eq:36}
\delta = -\left(\frac{l-1}{\rho}\right)\left(\frac{2}{3}\chi(l) \right) %
\left[-na^2e^{nat}+\frac{3a^2e^{2nat}}{(e^{nat}-1)^2}\right]
\end{equation}

\begin{equation}  \label{eq:37}
\gamma = \left(\frac{l+5}{2\rho}\right)\left(\frac{2}{3}\chi(l) \right) %
\left[-na^2e^{nat}+\frac{3a^2e^{2nat}}{(e^{nat}-1)^2}\right]
\end{equation}

\begin{equation}  \label{eq:38}
\eta = -\left(\frac{5l+1}{2\rho}\right)\left(\frac{2}{3}\chi(l) \right)\left[%
-na^2e^{nat}+\frac{3a^2e^{2nat}}{(e^{nat}-1)^2}\right]
\end{equation}

\begin{figure}[tbph]
\minipage{0.40\textwidth}
\centering
\includegraphics[width=\textwidth]{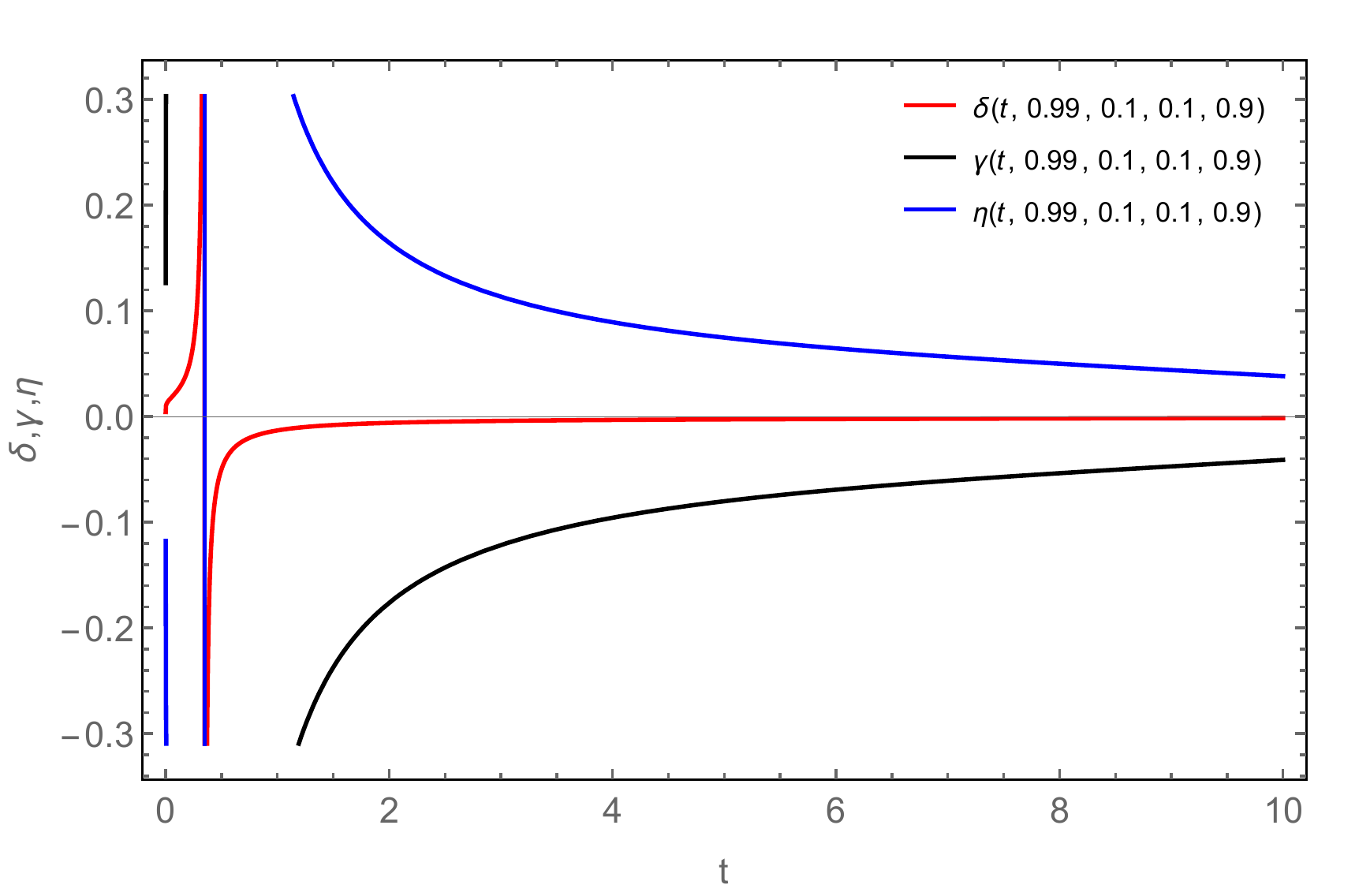}
\caption{Skewness parameters versus time $t$ \\
( $n=0.99,$ $a=0.1,$ $k=0.1,$ $l=0.9)$}
\endminipage
\minipage{0.40\textwidth}
\centering
\includegraphics[width=\textwidth]{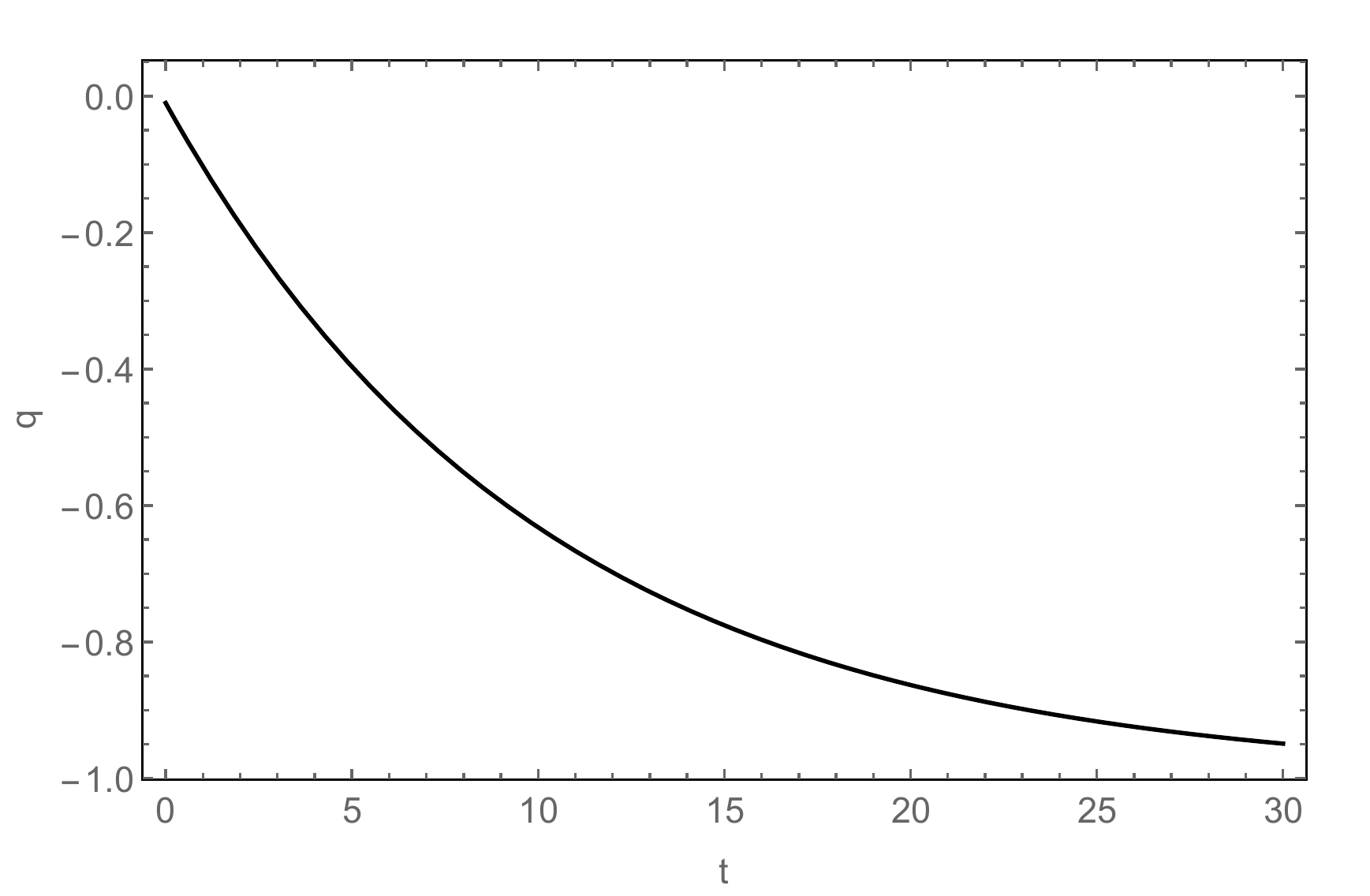}
\caption{Deceleration parameter $q$ versus time $t$\\
( $n=0.99,$ $a=0.1,$ $k=0.1,$ $l=0.9)$}
\endminipage
\end{figure}
FIG-11 describes the behaviour of the skewness parameters of the model. Here also the parameter $\delta$ aligned with the $x-$axis and  $\gamma$, $\eta$ remain as the mirror image of each other. In this case, the scalar expansion and shear scalar can be obtained
respectively as $\theta=\frac{3ae^{nat}}{e^{nat}-1}$ and $\sigma^2=\left(%
\frac{l-1}{l+1}\right)^2 \left(\frac{ae^{nat}}{e^{nat}-1}\right)^2$. Scalar expansion shows the change of volume of the corresponding universe in an accelerated way as it includes exponential terms. Similarly shear scalar indicated the deformation rate of the universe is positive always for $l>1$ and $a>0$.  The
deceleration parameter $q=-\frac{n}{e^{nat}(e^{nat}-1)}-(1-n)$.  The behaviour of deceleration parameter in FIG-12 indicates the acceleration expansion at late time; however at early time it might have decelerated. In order to get an accelerated expansion of the universe $n<1$. It can be noted that $q(t_0)\sim-0.7$ when $t_0=13.7$ billion years. 

\section{Conclusion}

We have considered three different functional forms of scale factor to
construct some DE cosmological models of the Universe in the
framework of GR. The space time at the backdrop considered
is the spatially homogeneous anisotropic Bianchi V metric. With the help of
the forms of scale factor and Hubble expansion rates along different
directions, the anisotropic behaviour of the dark energy driven cosmological
model has been simulated. In the first and third case, we have observed that at late
phase of cosmic dynamics, the accelerated expansion of the universe has
occurred; however before this the universe might have decelerating at early
time whereas in the second case it reveals the time period of cosmic transition from early deceleration phase to late time acceleration phase. The EoS parameters in case I and III become time varying. In case I, where the scale factor is in the form of general hybrid factor, the EoS parameters entirely remain in the quintessence region. However in case II, it starts evolving from phantom phase and in case III it evolves from phantom dominated era but approaches towards false vacuum model, lying primarily in the quintessence region. \\

The skewness parameters in case I and II indicates pressure anisotropy of late phase evolving dynamically with cosmic expansion. In all the cases skewness parameters evolve fluctuating from different ranges showing greater anisotropy. In case III, the parameters continue their behaviour towards an isotropic universe but in the other two cases pressure anirotropy remains constant. In the present work, we observed that, the constant in the functional form of the scale factors in \eqref{17} is used to reconstruct different DE anisotropic models, depending on Hubble parameter, we revealed more interesting results favouring phantom kind of behaviour and quintessence dominated era. It is concluded that pressure anisotropic in DE fluid play very important and interesting role of pressure anisotropic need to be carried out for further better understanding of the accelerated expansion of the universe.  

\section*{Acknowledgement}
BM and PPR acknowledge DST, New Delhi, India for providing facilities through DST-FIST lab, Department of Mathematics, where a part of this work was done. SKJP thanks NBHM, Department of Atomic Energy (DAE), Government of India for the post-doctoral fellowship. The authors are thankful to the anonymous referee for the valuable suggestions and comments for the improvement of the paper.

\end{document}